\documentclass[a4paper,12pt]{article}

\usepackage{amsfonts,amsthm,amsmath,amssymb,graphicx}
\usepackage{hyperref}
\usepackage{amsmath}
\usepackage{url}
\usepackage{graphicx,color}

 \def\a{{\alpha}}
 
 \def\frac#1#2{{#1\over #2}}

 \def\b{{\beta}}

\def\be{\begin{equation}}
\def\ee{\end{equation}}
\def\ba{\begin{eqnarray}}
\def\ea{\end{eqnarray}}
\numberwithin{equation}{section}

\date{}

\begin{document}

\title{Soldering freedom and BMS-like transformations}

\author{{Srijit Bhattacharjee}${}^{a}$, {Arpan Bhattacharyya}${}^{b,c}$\\
\it $^a$ Indian Institute of Information Technology, Allahabad,\\ 
Devghat, Jhalwa, Uttar Pradesh-211015, India\\
\it $^b$ Yukawa Institute for Theoretical Physics (YITP), Kyoto University,\\ Kitashirakawa Oiwakecho, Sakyo-ku, Kyoto 606-8502, Japan \\
\it  $^c$ Department of Physics and Center for Field Theory and Particle Physics,\\ Fudan University,
220 Handan Road, 200433 Shanghai, P. R. China\\}
\maketitle 

\abstract{When two spacetimes are stitched across a null-shell placed at the horizon of a black hole, BMS-supertranslation like soldering freedom arises if one demands the induced metric on the shell should remain invariant under the translations generated by the null generators of the shell. We revisit this phenomenon on the horizon of rotating shells and obtain BMS like symmetries. We further show that superrotation like soldering symmetries in the form of conformal isometries can emerge whenever the degenerate metric of any null hypersurface admits a dependency on null (degenerate direction) coordinate. This kind of conformal isometry can also appear for a null surface situated very close to the horizon of black holes. We also study the intrinsic properties of different kinds of horizon shells considered in this note. \tableofcontents

\newpage  
\section{Introduction}

Symmetry consideration has been a very powerful approach to study many physical systems, and spacetime geometry is not an exception. Many years ago Bondi, van der Burg, Metzner, and Sachs (BMS) studied the diffeomorphisms that preserve the asymptotic structure of an asymptotically flat spacetime at future null infinity $I^{+}$, and to their great surprise, the asymptotic symmetry group turned out to be infinite dimensional- a semidirect product of Lorentz group and {\it supertranslations} (angle dependent translations) \cite{BBM, Sachs,Newman1,Newman:1962cia,Barnich:2011ty}. These supertranslations constitute an infinite dimensional abelian subgroup of the BMS group and they map one asymptotically flat solution of Einstein's equation to another. Recently there has been a growing interest in determining the structure of asymptotic symmetries in gravity. Not long ago, BMS group has got extended and a new symmetry- called {\it superrotation} - has emerged at the null infinities (both future and past) of asymptotically flat spacetimes \cite{Barnich:2006av,Barnich:2009se,Barnich:2011ct,Barnich}\footnote{Recently this has been extended to Conformal BMS group in \cite{Haco}.}. Superrotations are understood as diffeomorphisms acting on the celestial spheres at $I^{\pm}$. In simple terms, this new symmetry rotates around each generator of asymptotic null infinities $I^{\pm}$ separately.  Both supertranslation and  superrotation like symmetries have also emerged from the study of diffeomorphisms that preserve the near horizon asymptotic structure of black holes \cite{Pinoetal,Pinoetal1,Shi1}. On the other hand, it has been shown that there is a deep connection between the infrared structure of gravity (and also some gauge theories) and the asymptotic symmetries of it. Certain subgroup of BMS$^+ \times $BMS$^-$ has emerged as an exact symmetry of the quantum gravitational S matrix \cite{Strominger:2013lka,Strominger:2013jfa, He:2014laa,Cachazo,He1,Lysov2, Campiglia1,Campiglia:2015qka,Kapec1,Dumitrescu:2015fej, Kapec3}. For a more recent and comprehensive review on this subject, interested readers are referred to  \cite{Strominger} and also encouraged to consult various references therein. These interconnections have generated the intriguing possibility of resolving black hole information paradox. The idea is: black holes are imparted by an infinite number of soft hairs corresponding to diffeomorphisms that act non-trivially on the phase space  of General Relativity, and these soft hairs (gravitons) would be responsible to restore the missing information of hard gravitons of Hawking radiation \cite{Hawking:2016msc,Hawking,Hawking1,Carney}. \footnote{Recently there are some works which suggest that the soft charges do not play any role in resolving the information paradox. Interested readers are referred to \cite{Mirbabayi,Bousso,Donnelly}.}\par
Supertranslation like transformations also arise in another context when one tries to solder two spacetimes across a thin null shell assumed to be situated at the event horizon of a black hole \cite{Blau}. This kind of shell is termed as horizon shell. It has been shown that there exists a considerable amount of freedom to solder two metrics across a horizon shell (can be situated at any Killing horizon) for which the induced metric remains invariant under the translations along the null generators.  In fact, the group of soldering transformations turns out to be infinite dimensional and a restricted class of it has the identical structure- like supertranslations in BMS group \cite{Blau, Koga}. The emergence of such symmetry is understood as the existence of a residual freedom (as long as the induced metrics from both sides of the shell match) to slide along the null generators those generate the horizon. \par
In \cite{Blau}, a detailed analysis is presented on finding the soldering group of Schwarzschild spacetime. In this note, we first extend the results of \cite{Blau} for the case of rotating spacetimes. We then discuss the intrinsic properties of the shell for generic soldering and BMS-supertranslation like symmetries. We adopt the intrinsic formulation and directly read-off the intrinsic quantities of the shell corresponding to the BMS like soldering transformations by comparing the {\it oblique} extrinsic curvature of two sides of the shell in a common intrinsic coordinate. This reduces some algebra compared to the off-shell extension method adopted in \cite{Blau}. This procedure helps us to reinterpret the results of \cite{Blau} in a slightly different way. Soldering of two metrics across a null surface is predominantly local construction and insensitive to the global or asymptotic structure of the ambient manifolds. This fact is apparent from our study of horizon shells in rotating BTZ spacetime. We find supertranslation like soldering freedom for BTZ shells also. The properties of the shells are also discussed. We also observe that the soldering freedom not only can be recovered for a horizon shell placed at the event horizon of a black hole but also for causal horizons like Rindler horizon or horizon in pure de Sitter space. \par
As extended BMS group contains the superrotation symmetry also, one may be interested to look for such kind of symmetries in the context of soldering transformations. However, for horizon shells situated at the black hole event horizon, there is no scope of finding such symmetries. The fundamental junction condition doesn't remain valid in such scenarios. Interestingly, in some cases where the horizon metric has non-trivial time dependence (eg. Penrose's cut-paste construction for Minkowski light-cone \cite{Penrose, Nutku-Penrose}), we know that a non-trivial soldering can be constructed by combining a shift of null-coordinate  $V$ with conformal isometries of the $2$-sphere (spatial part of null hypersurface). This set of conformal transformations can be related to the superrotations of asymptotic symmetry group \cite{Zhiboedov}. Here we have shown how conformal isometries (superrotation like symmetries) can also be accommodated within the similar framework in which supertranslation like soldering freedom emerges (demanding junction condition to remain invariant, or equivalently finding solution of Killing equation for the induced metric on the null surface). For any spacetime with constant scalar curvature, this kind of soldering will emerge. For spacetimes with dimension of spatial slice of the null surface $\leq 2,$ we will have an infinite dimensional group of soldering transformations arising because of infinite number of conformal isometries. In higher dimensions, we will only have a finite dimensional conformal group. Although this kind of superrotation like symmetries could not be retrieved at the horizon of a black hole but one can find it on a null surface situated slightly away from the black hole horizon. We have explicitly shown this near the horizon of a Schwarzschild black hole and studied the properties of such shells. 
\par
In Section~(2) we review the Israel junction condition briefly. In the next section emergence of BMS like transformations from soldering freedom is reviewed and how one can obtain the conformal isometries is discussed. To demonstrate the machinery - we have applied it to the horizons of Schwarzschild and Minkowski spacetimes. We use the intrinsic formulation to compute the conserved charges. Next section is devoted to find soldering freedom for rotating shells. We also obtain conserved quantities corresponding to soldering freedom for rotating shells.  In Section~(5), we consider the case where conformal isometries emerge due to soldering. We demonstrate the examples of such situations. We also show how conformal isometries can be recovered near to the horizon of a black hole. Finally, we conclude with discussions on our results and indicate future scopes.

\section{Brief review of Israel junction condition} \label{first}
Soldering of two spacetimes across a null hypersurface is a well-studied problem and much of our discussions will closely follow the works by Israel, Poisson etc. \cite{DI,Barrabes,Barrabes1, Poisson}. In this section, we start by briefly reviewing all the essential features of Israel junction conditions.  
Most commonly, in general relativity, the problem is to find the surface dynamics of a thin shell, where the surface is embedded in a spacetime ($\mathcal{M}$) of the form $\mathcal{M}=\mathcal{M}_{+}\cup \mathcal{M}_{-}.$ $\mathcal{M}_{-}$ and $\mathcal{M}_{+}$
 denote respectively the manifolds inside and outside of the shell together with the corresponding intrinsic metrics $g^{-}_{\mu\nu}(x_{-}^{\mu})$ and $g^{+}_{\mu\nu}(x_{+}^{\mu}).$  $x^{\mu}_{\pm}$ denote the coordinates of the manifolds $\mathcal{M}_{\pm}$. Both the manifolds have boundaries, with $\mathcal{M}_{+}$ and  $\mathcal{M}_-$ are defined to the future and past of null hypersurfaces $\Sigma_+$ and $\Sigma_-$ respectively. Further we consider a common coordinate system $x^{\mu}$, installed across the common boundary $\Sigma$ of two manifolds $\mathcal{M}_{\pm}$. This coordinate system overlaps with the coordinates $x^{\mu}_{\pm}$ in some open neighbourhoods of $\mathcal{M}_{\pm}$ containing $\Sigma$. These constructions are required only for presentational convenience as the junction conditions can be cast  independent of any coordinate system. Now suppose we define a set of intrinsic coordinates $\zeta^{a}$ on the surface of the shell $\Sigma$, across which the two manifolds will be joined. Also, $x^{\mu}|_{\Sigma}=\zeta^a$. Now we can project both $g^{+}_{\mu\nu}$ and $g^{-}_{\mu\nu}$ on the surface from both sides. Then the junction condition ensures the continuity of the metric induced from both sides: 
 \be \label{junc}
 g_{ab}= g^{+}_{\mu\nu}e^{+\mu}_{a}e^{+\nu}_{b}|_{\Sigma_+}=g^{-}_{\mu\nu}e^{-\mu}_{a}e^{-\nu}_{b}|_{\Sigma_-}.
 \ee
 Where, \[e^{\pm \mu}_{a}=\frac{\partial x^{\mu}_{\pm}}{\partial \zeta^a}\] are the tangent vectors to the surface. We will use the Greek alphabets for the spacetime indices and Latin ones for the hypersurface indices. The junction condition simply says that the hypersurfaces are isometric, i.e $\Sigma_{+}=\Sigma_{-}=\Sigma.$ Consequently (\ref{junc}) determines the  functional dependence of the coordinates (although not uniquely)  $x^{ \mu}_{\pm}$ on $\zeta^a.$ In the literature this junction condition is often written in the following form, 
 \be
 [g_{a b}]=g^{+}_{a b }|_{\Sigma_{+}}- g^{-}_{a b}|_{\Sigma_{-}}=0.
\ee
We introduced here the box $'[\, ]'$ notation which means for any tensor $A^{\mu}$,
\be
[A^{\mu}]= A^{\mu}|_{\Sigma_{+}}-A^{\mu}|_{\Sigma_{-}}.
\ee
In the rest of the paper, we will be only considering null shells.  We define the normal vectors $n_{\alpha}^{\pm}=\chi\partial_{\alpha}(\Phi(x^{\mu}_{\pm}))$  for both $\Sigma_{\pm}.$ $\chi$ is an arbitrary normalization. These normals are also generators for the null congruences orthogonal to both the hypersurfaces. The equations of the hypersurfaces are given by $\Phi(x^{\mu}_{\pm})=0$. We always work with future directing normal vectors for each side of the shell. To complete the basis we also have to define the auxiliary vector $N^{\pm}$ such that, 
\be \label{auxn} N.N|_{\pm}=0,n. N=-1|_{\pm}.\ee So together with $e^{\pm\mu}_a$ they form a complete basis. From the continuity of the null congruence it follows,
\be \label{junc1}
[n^{\mu}]=0=[N^{\mu}].
\ee
The normal vectors must satisfy $n.e_{a}|_{\pm}=0.$  Since same intrinsic coordinates $(\zeta^a)$ should be induced from $+$ and $-$ sides of the shell we must have, 

\be \label{junc2}
[e^{\mu}_{a}]=0.
\ee

This condition ensures the continuity of the null normal $n$ and the spacelike tangent vectors across the shell . Now we can use the distributional tensor calculus to derive the form of the stress tensor for thin shell such that the Einstein equations will be satisfied.\footnote{We have only quoted the important results. For detail derivations interested readers are referred to \cite{Barrabes,Barrabes1,Poisson}.}  Now in the common coordinate chart $\{x^{\mu} \}$ we express the metric covering both sides as a distribution valued tensor like,
\be \label{junc3}
g_{\mu\nu}=g^{+}_{\mu\nu}\mathcal{\theta}(\Phi)+g^{-}_{\mu\nu}\mathcal{\theta}(-\Phi),
\ee
where both $g^{+}$ and $g^{-}$ has been expressed in terms of the coordinates $x^{\mu}.$ Now if we compute the derivative of (\ref{junc3}),
\be \label{junc4}
\partial_{\alpha}g_{\mu\nu}=\partial_{\alpha}g^{+}_{\mu\nu}\mathcal{\theta}(\Phi)+\partial_{\alpha} g^{-}_{\mu\nu}\mathcal{\theta}(-\Phi)+[g_{\mu\nu}](\partial_{\alpha}\Phi)\delta(\Phi).
\ee
 Using (\ref{junc}) and (\ref{junc2})  the last term in (\ref{junc4}) becomes zero. So we end up with the following form for the Christoffel symbol,
\be
\Gamma^{\mu}_{\alpha\beta}=\Gamma^{+\mu}_{\alpha\beta}\mathcal{\theta}(\Phi)+\Gamma^{-\mu}_{\alpha\beta}\mathcal{\theta}(-\Phi).
\ee
 As a result the Riemann tensor takes the following form,
\be \label{junc5}
R^{\alpha}{}_{\beta\gamma\delta}= R^{+\alpha}{}_{\beta\gamma\delta}\mathcal{\theta}(\Phi)+R^{-\alpha}{}_{\beta\gamma\delta}\mathcal{\theta}(-\Phi)+\delta(\Phi) Q^{\alpha}{}_{\beta\gamma\delta},
\ee
where, $Q^{\alpha}{}_{\beta\gamma\delta}=-\Big([\Gamma^{\alpha}{}_{\beta\delta}]n_{\gamma}-[\Gamma^{\alpha}{}_{\beta\gamma}]n_{\delta}\Big).$
To satisfy the Einstein equation we start with the following form for the stress tensor, 
\be \label{junc6}
T_{\alpha\beta}=T^{+}_{\alpha\beta}\mathcal{\theta}(\Phi)+T^{-}_{\alpha\beta}\mathcal{\theta}(\Phi)+S_{\alpha\beta}\delta (\Phi),
\ee
where, 
\be \label{junc6b}
8\pi S_{\alpha\beta}=Q_{\alpha\beta}-\frac{1}{2}Q g_{\alpha\beta}.
\ee
(\ref{junc6b}) follows from the $\delta(\Phi)$ part of the Einstein equations. 
Then the stress tensor of the shell ($S_{ab}$) can be found by projecting (\ref{junc6b})  to the surface,
\be
S_{ab}=S_{\alpha\beta}e^{\alpha}_{a}e^{\beta}_{b}.
\ee
We further observe that because of the junction condition there is no discontinuity in the tangential derivatives of the metric, only there is a jump in the normal direction.
\be \label{junc7}
[\partial_{\alpha} g_{\mu\nu}]=-\gamma_{\mu\nu}n_{\alpha}.
\ee
From this we get,
\be
\gamma_{ab}=N^{\alpha}[\partial_{\alpha} g_{ab}]=2[\mathcal{K}_{ab}].
\ee
$\mathcal{K}_{ab}$ is known as the `{\it transverse}' or {\it `oblique' } extrinsic curvature. This is defined as,
\be
\mathcal{K}_{ab}=e^{\alpha}_{a}e^{\beta}_{b} \nabla_{\alpha}N_{\beta}
\label{ecdef}.\ee

 Projecting $\gamma_{\mu\nu}$ to the surface gives us a unique induced metric on the surface of the shell. Using this we can finally recast $S_{\alpha\beta}$ (after some manipulation )\cite{Barrabes,Barrabes1,Poisson} in the following form,

\be 
S^{\alpha\beta}=\mu n^{\alpha}n^{\beta}+j^{A}(n^{\alpha}e^{\beta}_{A}+e^{\alpha}_{A}n^{\beta})+p  \sigma^{AB}e^{\alpha}_{A}e^{\beta}_{B}, \label{ST}\ee
where  $\sigma_{AB}$ is the non degenerate metric of the spatial slice of the surface of the null shell. Capital indices $A,B$ denote the spatial indices of the null surface. $\mu, J^{A}$ and $p$ can be interpreted as the surface energy density, current and pressure of shell respectively.  For the null shell these quantities can be related to the jump of the extrinsic curvature in the direction transverse to the shell ({\it oblique extrinsic} curvature). To make things concrete, throughout the paper we work in Kruskal coordinates. On the surface  $\Sigma$ of the null shell we define a coordinate chart such that $\zeta^{a}=\{V,x^{A}\}.$ $V$ is the parameter along the hypersurface generating null congruences.  In this coordinate system the normal vector takes the following form, 
\be \label{non1} n^{\alpha}=(\partial_{V})^{\alpha}.\ee
On $\Sigma$ the metric will take the form $g_{ab}d\zeta^ad\zeta^b= g_{AB} dx^Adx^B.$ One then adopts a single chart $x=\{ U, x^{a}\}$ such that $x^{a}|_{\Sigma}=\zeta^{a}.$ Given this coordinate chart the conditions (\ref{junc}), (\ref{junc1}) and (\ref{junc2}) are satisfied. Equipped with this coordinate system, we write down the conserved quantities in the following way \cite{Barrabes,Barrabes1, Poisson},

\begin{align}
\begin{split} \label{junc8}
\mu=-\frac{1}{8 \pi}\sigma^{AB}[\mathcal{K}_{AB}],\, J^{A}=\frac{1}{8 \pi}\sigma^{AB}[\mathcal{K}_{VB}],\,
p=-\frac{1}{8 \pi}[\mathcal{K}_{VV}],\end{split}
\end{align}

where, $[\mathcal{K}_{VV}]=\frac{1}{2}\gamma_{\alpha\beta}n^{\alpha}n^{\beta},\,[\mathcal{K}_{VA}]=\frac{1}{2}\gamma_{\alpha\beta}e^{\alpha}_{A}n^{\beta},\, [\mathcal{K}_{AB}]=\frac{1}{2}\gamma_{\alpha\beta}e^{\alpha}_{A}e^{\beta}_{B}$. The shell's  stress-energy tensor obeys certain conservation equations. The detailed derivation can be found in \cite{Barrabes}. We display here the most relevant one for our analysis. 

\be 
N^a(\partial_b + \tilde{\Gamma}_b)S^{ab}-S^{ab}\tilde{\mathcal{K}}_{ab}=0 \label{consv}.
\ee
Here $\tilde{\Gamma}_b$ and $\tilde{\mathcal{K}}_{ab}$ denotes arithmetic mean of $\Gamma^{\pm \mu}_{\mu b}$ and $\mathcal{K}^{\pm}_{ab}$ respectively. 

\section{Soldering freedom and BMS-like transformations}  \label{third}
It is well known that there exists a considerable amount of freedom in the choice of intrinsic coordinate on $\Sigma$  in the  null direction. This is often termed as ``soldering freedom."  Recently in \cite{Blau}, this fact has been utilized to show that BMS (supertranslation) type transformations emerge on the horizon-shell. In this section, we will briefly review the essential points of this development and study the soldering symmetries for rotating spacetimes. 

We can ask what are the possible allowed coordinate transformations on either side of the shell preserving the junction condition (\ref{junc}). In a suitable coordinate system this boils down to solve for the Killing vectors ($Z^a$)  of $g_{ab},$ the metric on $\Sigma.$ So we solve the following equations,  
\be \label{killing}
\mathcal{L}_{Z} g_{ab}=0.
\ee
From (\ref{killing}) we get,
\be \label{killing1}
Z^{c}\partial_{c}g_{ab}+(\partial_{a}Z^{c})g_{cb}+(\partial_{b} Z^{c})g_{ca}=0.
\ee
In \cite{Blau} the authors have worked with metrics where $g_{aV}=0.$ From the $a\,V$ components of (\ref{killing1}) we can easily conclude ,
\be
\partial_{V} Z^{A}=0.
\ee
 This implies that the isometry transformations along the spatial directions of null surface are independent of $V$. Next if we consider the spatial components $ A\, B$  of (\ref{killing1}) we get,
\be \label{killing2}
Z^{V}\partial_{V} g_{AB}+ Z^{C}\partial_{C}g_{AB}+(\partial_{A}Z^{C})g_{CB}+(\partial_{B}
Z^{C}) g_{C A}=0.
\ee
Now there are two possibilities here. First one is the case where the metric $g_{AB}$ is independent of $V,$  and we are free to choose $Z^{V}$ as an arbitrary function of $(V,x^{A}).$
\be
Z^{V}= F (V, X^{A}).
\ee
For Killing horizon one can investigate the effect of this isometry transformation on the normal vector $ n^{a}.$ One can show the isometry transformations preserve the direction of the normal to the shell \cite{Blau}, this implies Lie transport of the normal vector should be proportional to itself i.e. $\mathcal{L}_{Z}n^{a} \sim n^a.$ If one considers only those transformations for which $\mathcal{L}_{Z} n^a=0$, then it gives,
\be \label{killing3}
\partial_{V} Z^{V}=0.
\ee
(\ref{killing3}) implies,
\be\label{killing4}
Z^{V}= F(x^A).
\ee
 From this relation it is apparent that $Z^V$ generates the following transformation,
\be
V\rightarrow  V+ F (X^A).
\ee
One can immediately identify this with the supertranslation, already known in the literature. In \cite{Blau} the shell stress tensor evaluated by extending the generator $Z$ off the shell. But here we will take a slightly different approach namely the intrinsic formalism following \cite{Barrabes,Barrabes1,Poisson}. This will help us to visualize the emergence of BMS like symmetries as soldering freedom of two manifolds across a null surface in a different way: we always have a seed metric (like Schwarzschild) and the horizon shell sort of divides the spacetime into two manifolds. Now, if one performs supertranslation like coordinate transformations in one side (say $+$ side of the hypersurface) of the horizon shell and stitches that with the other side ($-$) without altering the fundamental junction condition, then the generators of those supertranslations also preserve the induced metric on the hypersurface. We then calculate the oblique extrinsic curvatures of the two sides of the shell and construct the stress tensor of the shell. Thus we can obtain the conserved quantities with this construction quite easily. To appreciate this fact let us consider the Schwarzschild case as depicted in \cite{Blau}. We consider a `supertranslated' Schwarzschild as $\mathcal{M}_{+}$ and the metric of  $\mathcal{M}_{-}$ side remains as the seed Schwarzschild metric. Recall supertranslation-like soldering freedom which is \[V\rightarrow  V+ F (X^A).\]

Following is the supertranslated metric in the $+$ side, written in Kruskal coordinates:

\be
ds^2= -G(r) (dU_+dV_+ +\partial_{\theta_+}F(\theta_+,\phi_+)dU_+d\theta_+ +\partial_{\phi_+}F(\theta_+,\phi_+)dU_+d\phi_+)+r^2(U_+,V_+)d\Omega_2^2,
\ee
where \be G(r)=\frac{32m^3}{r} e^{-r/2m}. \ee

The metric on the other side is the usual Schwarzschild metric and we assume they coincide with the intrinsic coordinates of the shell. The horizon is situated at $U=0$ surface. Equipped with this we can compute the conserved quantities from the jump of the extrinsic curvature (\ref{junc8}). Recall Eq. (\ref{ecdef}), this is used to compute the ``{\it oblique}" extrinsic curvature. The auxiliary normal, satisfying (\ref{auxn}) is given by:

\be \label{non2} N_{\alpha}=-(\partial_V)_{\alpha}.\ee

As already mentioned, the oblique extrinsic curvatures are computed for both sides $(\pm)$ of the shell in a common coordinate system. Only $[\mathcal{K}_{\theta\theta}]$ and $[\mathcal{K}_{\phi\phi}]$ are non-vanishing. Thus we get,
\be \label{en}
\mu=-\frac{1}{8\pi}\sigma^{AB}[\mathcal{K}_{AB}]=-\frac{1}{32 m^2 \pi}\Big(\Delta^{(2)} F -2F\Big).
\ee
The shell conserved energy now is obtained by integrating $\mu$ on the spatial slice of the horizon.
\be
E[F]=\frac{1}{8\pi }\int_{S^2} F(\theta, \phi),
\ee
This reproduces the charge corresponding to the supertranslation on the horizon as described in Eq. (6.31) of \cite{Blau}. The properties of such horizon-shells are studied in detail in \cite{Blau}. Specially it was concluded that there will not be impulsive gravitational wave without any matter on the shell.
 However, this situation alters when one considers Minkowski light cone.

\subsection*{Minkowski  spacetime}
Let us consider 4-dimensional Minkowski spacetime. Usually we write it in terms of Cartesian coordinates in the following way. 

\be
ds^2=dx^2-dt^2+dy^2+dz^2.
\ee
Next we define,
\be
U=t-x, V=t+x.
\ee
The metric becomes,
\be \label{min1}
ds^2=-dU dV+dy^2+dz^2.
\ee
Let us now consider the null surface (``light-front") defined by $U=0$ and solder two Minkowski spaces  across this surface.  This set up produces plane-fronted lightlike signal. As discussed previously we can immediately check this construction generates soldering freedom of the form,
\be
V\rightarrow F(V,y,z).
\ee
After obtaining this soldering transformation we can compute the stress tensor for the shell using (\ref{junc8}). 
\begin{align}
\begin{split}
p=\frac{1}{4\pi}\frac{\partial_{V}^2 F}{\partial_{V} F}, J^{A}=-\frac{1}{4\pi}\frac{\partial_{A}\partial_{V}F}{\partial_{V}F},\mu=\frac{1}{4\pi}\frac{\Delta^{(2)} F}{\partial_V F}.
\end{split}
\end{align}
Then we demand $p=0,J^{A}=0.$ This gives upto a rescaling factor,
\be  \label{mit}
F(V,y,z)=V+T(y,z).
\ee
This is precisely a supertranslation type transformation. One can express the energy density as,
\be \label{en3}
\mu=\frac{1}{4\pi}\Delta^{(2)} T.
\ee
If we now demand that the entire stress tensor of the shell should vanish we get,
\be
\mu=0.
\ee
From this we get,
\be
\Delta^{(2)} T=0.
\ee
Now this admits a regular solution. We get,
\be \label{psol}
T(y,z)= c_1\, f(y+i\,z)+c_2\, f(y-i\,z).
\ee
It's a linear combination of two arbitrary functions of $y\pm i\, z.$ This case admits regular solution for $T(y,z)$ unlike the cases of Killing horizons of black holes.  So, if we join shells with non-compact topology like a plane, in this case, we get shells supporting impulsive light like signals without any matter \cite{Barrabes1}. One can now have many interesting cases. For example one can have $T(y,z)=\frac{a}{2}[(y+iz)^2+(y-iz)^2]+{b \over 2}[(y+iz)^2-(y-iz)^2]$, for which one gets pure impulsive gravitational wave.\footnote{One can also compute Newman-Penrose scalar $\Psi_4$ and find if it is non-zero. A non zero $\Psi_4$ indicates existence of gravitational wave.}  On the other hand if one considers a term :$-\frac{c}{2}(y^2+z^2)$, in addition to the above soldering terms, a non-zero matter density $\mu=c/4\pi$ in addition to the impulsive gravitational wave appears. Therefore the matter and the gravity wave can co-exist in this scenario, a known fact from the earlier studies of singular null hypersurfaces \cite{Barrabes1}. 
\par 
\begin{itemize}
\item{ {\bf  Rindler horizon}:\\
We know that an accelerated observer in Minkowski spacetime sees a horizon in front of it. The spacetime seen by the observer can be captured by performing the following transformations:
\be
U=-\frac{1}{a} e^{-a\, u},\quad V=\frac{1}{a} e^{a\, v},
\ee
where  $a$ is the proper acceleration of the Rindler observer. For this case, we get a supertranslation like soldering freedom related to (\ref{mit}) as follows,
\be
T(x,y)=\frac{1}{a}e^{a \tilde T(x,y)}.
\ee
Subsequently we get, 
\be
\mu=\frac{e^{a \tilde T(x,y)}}{4\pi}\Big( \Delta^{(2)} \tilde T+a\,\{ (\partial_x \tilde T)^2+(\partial_{y} \tilde T)^2\}\Big).
\ee

To examine the fact if we get a situation where gravitational wave exists without any matter on the shell, we again demand $\mu=0$ and can find a solution of the type
\be
\tilde T(x,y)=\frac{1}{a} \log[c(x\pm y)].
\ee}
This indicates the existence of lightlike signal through the accelerated horizon of Rindler spacetime.

Infinite dimensional soldering symmetry can also be attributed to the cosmological horizon of de-Sitter space which is an example of Killing horizon but not an event horizon. However this horizon is genuine unlike the Rindler horizon, which only exists for an accelerated observer.

\item{{\bf Spacetime describing infinite straight light like string}\\  We start with a spacetime posessing a cosmic string and then give it infinite boost. The limiting geometry takes the following form,
\be
ds^2=-dU dV+dy^2+ dz^2-8\pi \mu |y| \delta(U) dU^2.
\ee 
$U=0$ is a null surface.  We can easily verify the following supertranslation like soldering freedom should appear across this surface, $ V\rightarrow V+ T(y,z)$. Also demanding that the shell's stress tensor vanishes we again get a solution for $T(y,z)$ similar to  (\ref{psol}). }
\end{itemize}


\subsection{Soldering freedom redux}\label{redux}
We now look back at (\ref{killing2}) and explore the second possible scenario. Till now we have discussed the case where $g_{AB}$ is independent of $V.$ Generically, when $g_{AB}$ depends on $V, $ this equation implies $Z^{V}=0$ and there is no nontrivial soldering freedom. Now we will examine such cases where $g_{AB}$ indeed depends on $V$ but we will still have nontrivial soldering freedom. First we allow conformal transformations of the spatial slice such that,
\be \label{conf}
Z^{C}\partial_{C}g_{AB}+(\partial_{A}Z^{C})g_{CB}+(\partial_{B}
Z^{C}) g_{C A}=\Omega(x^{A})g_{AB}.
\ee
This is compatible with (\ref{killing1}) as $\Omega(x^{A})$ is independent of $V$ but can be in general any function of $x^A$ only to be constrained by  (\ref{conf}). Then from (\ref{killing2}) we have,
\be
Z^{V}\partial_{V}g_{AB}+\Omega(x^{A}) g_{AB}=0.
\label{zv}\ee
 Next we explore feasible solution of the above equation. Let us consider the following situation,
 \be
 g_{AB}=V^2 \tilde g_{AB},
 \label{conf-decomp}\ee
 where $\tilde g_{AB}$ is only a function of $x^{A}.$ This may appear to be pretty non-generic, but (\ref{conf-decomp}) seems to provide a feasible solution for $Z^V$ in Eq. (\ref{zv}). In fact one can consider any regular function of $V$ instead of $V^2$. That will provide us with the most general situation. The plausibility of above assertion can be argued as follows: Without loss of much generality we can assume that we can install Gaussian null coordinates (eg. Kruskal type) near the horizon of spacetimes that are of our interest. Then the tensorial equation (\ref{zv}) yields the following condition between the diagonal components $g_{11}=f_1(V,\a,\b),\,g_{22}=f_2(V,\a,\b)$ of the $2-$metric $g_{AB}$,
 \ba
 \partial_V \ln f_1(V,\a, \b)&=&\partial_V \ln f_2(V,\a,\b) \nonumber \\
 \frac{f_1(V,\alpha,\beta)}{f_2(V,\alpha,\beta)}&=&C(\alpha, \beta).
 \label{cndtnV}
 \ea
 Where $\alpha, \beta$ are coordinates on the $2-$surface and $C(\a,\b)$ is the integration constant. Clearly (\ref{cndtnV}) is always satisfied if both $f_1$ and $f_2$ can be cast as product of a regular function of $V$, $g(V)$ and some function (not necessarily the same) of $\a,\b$. 
 
 Now from (\ref{conf-decomp}) we have,
 \be \label{mint1a}
 2\, V\, Z^V\tilde g_{AB}+\Omega(x^{A}) V^2 \tilde g_{AB}=0.
 \ee
 This gives for all components of $\tilde g_{AB}$ ,
 \be \label{mint2}
 Z^{V}=-\frac{V\,\Omega(x^{A})}{2}.
 \ee
 When the dimension of the spatial slice is greater than 2 then the functional form of the  $\Omega(x^{A})$ is completely fixed and number of generators of such symmetry is also finite. So we  get a finite dimensional soldering group with $Z^{V}$ completely fixed. When the dimension of the spatial slice is $2$ then $\Omega(x^{A})$ can in principle be unconstrained and hence the soldering group will also be infinite dimensional. We will demonstrate this by explicit constructions in Section~(5). 

\section{Soldering freedom and BMS-like transformations for rotating shells}
We now  generalize the constructions depicted in \cite{Blau} for metrics where $g_{a V} \neq 0$ (expressed in some Gaussian null coordinates).  We take two specific examples : Kerr spacetime in slow rotation limit and rotating BTZ. As before, we will use the coordinates of $\mathcal{M}_{-}$ as the intrinsic coordinates covering both sides, $x^{\alpha}_-=x^\alpha.$ Also $ x_{-}^{\alpha}|_{\Sigma}= \zeta^a.$ The horizon in the Kruskal coordinates is identified by setting $U=0$ for both $\mathcal{M}_{\pm}.$  These two spacetimes will be isometrically soldered at $U=U_+=0.$ The components $g_{aV}$ is proportional to $U$ such that at $U=0,$ $g_{aV}$ will go to zero. So it can be easily checked that (\ref{killing1}), (\ref{killing2}) and (\ref{killing3}) remain the same.  On the horizon, we have $Z^{V}=F(V,X^B),$ which corresponds to the soldering freedom along the null direction. This will again produce the supertranslation like transformations. We then compute the shell stress tensor using  (\ref{ST}) and (\ref{junc8}).
\subsection*{\begin{itemize} \item Horizon shell in slowly rotating Kerr spacetime \end{itemize}}
We consider first the  Kerr metric in slow rotation limit i.e. we shall work up to first order in rotation parameter. The virtue of the slow rotation limit is, we will have more analytic control. In Kruskal coordinates, the metric takes the following form, 
\begin{align}\begin{split} \label{kerr1}
ds^2=&r^2(d\theta^2+\sin(\theta)^2d\tilde \phi^2)-\Big(\frac{32m^3}{r}\Big)e^{-r/2m}dUdV\\&+\frac{2a}{r}\sin(\theta)^2e^{-r/2m}(r^2+2 m r+4 m^2) d\tilde \phi (U dV-V dU).
\end{split}
\end{align}$a$ is the rotation parameter and we expand all the quantities in small $a$ and keep only those  terms which are linear in $a$. 
 On  the horizon ($\Sigma$) the induced metric takes the following form,
\be
ds^2|_{\mathcal{N}}=r^2(d\theta^2+\sin(\theta)^2d\tilde \phi^2)|_{\mathcal{N}}.
\ee
As mentioned before we take $\{U_{-}, V_{-},\theta_{-},\tilde \phi_{-}\}$ as the intrinsic coordinates,  $\{U, V,\theta, \tilde \phi\}.$  Once again we will have $Z^{V}= F(V,\theta,\tilde \phi)$  upon solving the Killing equations on the horizon. The angular components of the killing equations produce the usual isometry transformations on the 2-sphere. We then do an active coordinate transformation of the form $V\rightarrow V+F(V,\theta,\tilde \phi)$ on the  $-$ side  and solder this with usual Kerr metric on $+$ side. Then using (\ref{junc8}), and imposing the condition $\mathcal{L}_{Z} n^a=0$,  we can write down the intrinsic energy momentum tensor of the shell from the jump of the oblique extrinsic curvature. The soldering transformation now becomes supertranlsation like.
\be
F=a V+B(\theta,\tilde \phi),
\ee
where, $a$ is a scale factor and $B(\theta,\tilde \phi)$ is an arbitrary function of angular coordinates. 
So in general upto a rescaling factor we have,
\be
F(V,\theta,\tilde \phi)= V+ T(\theta,\tilde \phi).
\ee

\subsection*{\begin{itemize}\item Off-shell extension of soldering transformations for Kerr spacetime\end{itemize}}
The intrinsic properties of the horizon shells for generic soldering transformation can be read off by extending the generators of soldering freedom off the shell. This method was advocated in \cite{Blau}. Here we demonstrate the method for slowly rotating Kerr and we can check that this matches with the results obtained from the jump in the relevant components of the {\it oblique} extrinsic curvature method as discussed in Section~(\ref{first}). So this way of finding the conserved quantities will serve as a complementary method to the one discussed in Section~(\ref{first}). We need to impose the following conditions for extending the generators off the shell say in $(+)$ side,
\be
L_{Z_{+}}g_{\alpha \beta}|_{\Sigma}=0.
\label{ofshell}\ee
Where $Z_+$  is the off shell extension of soldering symmetry generator $Z=Z^V\partial_V$. In Kruskal type coordinates we write this vector as,
\be
Z_+=Z^V\partial_V + Uz^{\alpha}\partial_{\alpha}.
\ee

Lifting the soldering symmetry off the shell means we have on the shell: $ Z_+|_{\Sigma}=Z^V\partial_V$ and $\partial_a Z_+|_{\Sigma}=(\partial_aZ^V)\partial_V$. In our adopted coordinates $g_{aV}|_{\Sigma}=0=\partial_Vg_{ab}|_{\Sigma}$, and the relation 

\be
L_{Z_+} g_{ab}|_{\Sigma}=0
\ee
is identically satisfied. Hence the condition (\ref{ofshell}) reduces to 

\be
L_{Z_{+}}g_{U \beta}|_{\mathcal{\Sigma}}=0.
\label{ofshell1}\ee

Below we write down all the components explicitly (all of these equations are satisfied on the shell). 
\begin{align}
\begin{split} \label{kerr2}
&Z^{V}\partial_{V}g_{UU}+2 z^{\tilde\phi}g_{\tilde \phi U}+ 2 z^{V}g_{VU}=0,\\&Z^{V}\partial_{V}g_{UV}+\Big(z^{U}+(\partial_{V}Z^{V})\Big)g_{UV}=0,\\&Z^{V}\partial_{V}g_{UA}+z^{\alpha}g_{\alpha A}+(\partial_{A} Z^{V})g_{UV}=0.
\end{split}
\end{align}
Solving (\ref{killing1}) upto $\mathcal{O}(a)$,
\begin{align}
\begin{split}
&z^{V}=-z^{\tilde\phi}\frac{g_{\tilde \phi U}}{g_{VU}}=-z^{\tilde \phi}\frac{3 a V }{2 m}\sin(\theta)^2=-(\partial_{\tilde \phi} Z^{V})\frac{3 a V}{e m},\\&z^{U}=-\partial_{V} Z^{V},\\&z^{\tilde\phi}=\Big(Z^{V} -(\partial_{V} Z^{V})V\Big)\frac{3 a}{e m}+(\partial_{\tilde \phi} Z^{V})\frac{2}{e \sin(\theta)^2},\\&z^{\theta}=\frac{2}{e}(\partial_{\theta} Z^{V}).
\end{split}
\end{align}
Then using $Z^{V}= V\omega(V,\theta,\tilde \phi)$ we get the required Killing vector for off-shell transformation,
\begin{align}
\begin{split} \label{kerr3}
Z_{+}&=\omega (V \partial_{V}-U \partial_{U}) +U V \Big(\frac{2}{e \sin(\theta)^2} (\partial_{\tilde\phi} \omega)\partial_{\tilde\phi}+\frac{2}{e}(\partial_{\theta}\omega)\partial_{\theta} -(\partial_{V}\omega)\partial_{U}\Big)\\&-\frac{ 3 a \,U V^2}{e m} \Big(( \partial_{\tilde \phi} \omega)\partial_{V} +  (\partial_{V}\omega)  \partial_{\tilde\phi} \Big).
\end{split}
\end{align}
We now consider the following ansatz to obtain the finite counterparts of the infinitesimal transformations given in (\ref{kerr3}),
\begin{align}
\begin{split} \label{kerr4}
V_{+}&=F(V,\theta, \tilde \phi)+U A(V,\theta,\tilde \phi),U_{+}=U C(V,\theta,\tilde \phi),\\& \theta_{+}=\theta+U B^{\theta}(V,\theta,\tilde \phi),\tilde \phi_{+}=\tilde \phi+U B^{\tilde \phi} (V,\theta,\tilde \phi).
\end{split}
\end{align}

Demanding the continuity of full spacetime metric across the junction at leading order in $U$ we get ( retaining terms upto $\mathcal{O}(a)$),
 \begin{align}
\begin{split} \label{kerr5}
&C=\frac{1}{\partial_V F}\,\\&
B^{\theta}=\frac{2}{e} \frac{\partial_{\theta}F}{\partial_V F}\,\\&
B^{\tilde \Phi}= \frac{2}{e}\frac{1}{\sin(\theta)^2}\frac{\partial_{\tilde \phi}F}{\partial_{V} F}+\frac{3a}{2e\,m }(\frac{F}{\partial_{V} F}- V)\,\\&
A=\frac{e}{4}\partial_{V} F \Big((\frac{2}{e} \frac{\partial_{\theta}F}{\partial_V F})^2+\sin(\theta)^2(\frac{2}{e}\frac{1}{\sin(\theta)^2}\frac{\partial_{\tilde \phi}F}{\partial_{V} F})^2\Big)-\frac{3 a\, \partial_{\tilde \phi}F \, V}{ 2e m }.
\end{split}
\end{align}
Now we evaluate $\gamma_{ab}$ and the conserved charge as defined in (\ref{junc7}), and (\ref{junc8}).  On $\mathcal{M}_{-}$ we have,
\be
r(UV)^2=4 m^2- \frac{8 m^2}{e} U V+\cdots.
\ee
and on $\mathcal{M}_{+}$ we have,
\be
r(U_{+} V_{+})^2=4 m^2- \frac{8 m^2}{e} U_{+} V_{+}+\cdots= 4m^2-\frac{8 m^2}{e} \frac{U F}{\partial_{v} F }.
\ee
Also,
\be
\sin(\theta_{+})^2=\sin(\theta)^2+2 U \Theta\sin(\theta)\cos(\theta)+\cdots.
\ee

We now expand the tangential components of the metrics in $\mathcal{M}_{+}$ and $\mathcal{M}_{-}$ to linear order in $U$. For $\mathcal{M}_{-}$  we have,

\begin{align}
\begin{split}
g^{-}_{ab}dx^{a} dx^{b}=g^{0}_{AB} dx^{A} dx^{A}- U \frac{8 m^2}{e} \Big( -\frac{3\, a}{2 m} \sin(\theta)^2 dV d\tilde \phi+V d\theta^2+ V \sin(\theta)^2 d \tilde \phi^2\Big).
\end{split}
\end{align}
For $\mathcal{M}_{+}$ we have, 
\begin{align}
\begin{split}
g^{+}_{ab}dx^{a} dx^{b}=g^{0}_{AB} dx^{A} dx^{B}+8 m^2 U &\Big(-\frac{2}{e}dA dF+\sigma_{AB} dx^A(dB^B-(F/eF_V) dx^{B})+\\& \sin(\theta)\cos(\theta)\,\Theta\,\, d\tilde\phi^2+\frac{6 \, a \,\sin(\theta)^2}{4 e m } d\tilde \phi(AdF-F dA) \Big).
\end{split}
\end{align}
Now,
\be
\gamma_{ab}=N^{\alpha}[\partial_{\alpha} g_{ab}]= N^{U}[\partial_{U} g_{ab}]+N^{\phi}[\partial_{\tilde \phi} g_{ab}]
\ee

and 
\be
N^{U}=\frac{e}{8 m^2}\,, N^{\tilde \phi}=\frac{3 a V}{16  m^3}.
\ee
Using (\ref{kerr5}) we get, 
on $\mathcal{M}_{+}$ , 
\begin{align}
\begin{split}
g^{+}_{ab}dx^{a} dx^{b}=g^{0}_{AB} dx^{A} dx^{B}+8 m^2 U &\Big(\frac{2}{e}\Big( \frac{\partial_{V}\partial_{a} F}{\partial_{V} F }dV dx^a\Big)+\frac{2}{e}\frac{\partial_{A}\partial_{B} F}{\partial_{V} F}dx^{A} dx^{B}-\frac{4}{e}\cot(\theta)\frac{\partial_{\tilde \phi} F}{\partial_{V} F} d\theta d \tilde \phi\\&-\sigma_{AB} dx^A(F/eF_V) dx^{B}+ \frac{2}{e}\sin(\theta)\cos(\theta)\,\frac{\partial_{\theta} F}{\partial_{V} F}  d\tilde \phi^2\\&+\frac{3a}{2em}\sin(\theta)^2 d\tilde \phi\Big(2\frac{\partial_{B} F}{\partial_{V} F}dx^{B} +dV \Big)\Big).\end{split}
\end{align}
For $\mathcal{M}_{-}$ we have,
\begin{align}
\begin{split}
g^{-}_{ab}dx^{a} dx^{b}=g^{0}_{AB} dx^{A} dx^{B}- U \frac{8 m^2}{e} \Big( -\frac{3\, a}{2 m} \sin(\theta)^2 dV d\tilde \phi+V d\theta^2+ V \sin(\theta)^2 d \tilde \phi^2\Big).
\end{split}
\end{align}
So to $\mathcal{O}(a)$ we have,
\begin{align}
\begin{split} \label{kerr6}
&\gamma_{V a }=2\frac{\partial_{V}\partial_{a} F}{\partial_{V} F },\\&\gamma_{\theta \theta }= 2\Big(\frac{\nabla_{\theta}^{(2)} \partial_{\theta} F}{\partial_{V} F }-\frac{1}{2}\Big(\frac{F}{\partial_{V} F}-V\Big)\Big),\\&\gamma_{\theta \tilde \phi }= 2\Big(\frac{\nabla_{\theta}^{(2)} \partial_{\tilde \phi} F}{\partial_{V} F  }+\frac{3 a\sin(\theta)^2}{2 m}\frac{\partial_{\theta} F}{\partial_{V} F}\Big),\\&\gamma_{\tilde \phi \tilde \phi }= 2\Big(\frac{\nabla_{\tilde\phi}^{(2)} \partial_{\tilde \phi} F}{\partial_{V} F }-\frac{1}{2}\sin(\theta)^2\Big(\frac{F}{\partial_{V} F}-V\Big)+\frac{3 a \sin(\theta)^2 }{ 2m }\frac{\partial_{\tilde \phi} F}{\partial_{V} F}\Big).\end{split}
\end{align}

Using (\ref{kerr6}) we can write down the intrinsic energy momentum tensor off the shell. From that we get, 
\begin{align}
\begin{split} \label{kerr7}
&p=-\frac{1}{16 \pi} \gamma_{VV}=-\frac{1}{8\pi}\frac{\partial_{V}^2 F}{\partial_{V} F},\\&j^{A}=\frac{1}{32 m^2 \pi}\sigma^{AB}\frac{\partial_{B}\partial_{V}F}{\partial_{V} F},\\&\mu =-\frac{1}{32 m^2 \pi \partial_{V} F}\Big(\nabla^{(2)} F-F +V\partial_{V}F  +\frac{3 a}{2m}\partial_{\phi} F\Big).
\end{split}
\end{align}
The energy density of the shell is now takes the following form,
\be
\mu=-\frac{1}{32 m^2 \pi} \Big(\nabla^{(2)} T-T+\frac{3 a}{2m}\partial_{\tilde \phi} T \Big),
\ee
which implies,
\be
\partial_{V} \mu=0.
\ee
The total energy is given by integrating $\mu$ over the spacelike cross-section of the horizon.
\be \label{kerr7a}
E= \frac{1}{32m^2 \pi}\int d\theta d\tilde \phi \sqrt{g}\,T,\ee
where, $\sqrt{g}$ is the determinant of the induced metric of the spacelike cross-section of the horizon of the null shell. 

\vskip1cm
\newpage 
{\bf Next  we consider several special situations: }
\begin{itemize}
\item { Zero pressure, $p =0:$  We get,   $\partial_{V}^2 F=0.$  This in turn implies $ F= f(\theta,\phi) V+ g(\theta, \phi)$. This transformation as usual leaves surface gravity invariant.} 
\item{ Both pressure and current zero, $p=0, J^{A}=0: $}
\end{itemize}

This gives us, 
\be
F=a V+B(\theta,\tilde \phi),
\ee
where, $a$ is a scale factor and $B(\theta,\tilde \phi)$ is an arbitrary function of angular coordinate. 
So in general upto a rescaling factor we have,
\be
F(V,\theta,\tilde \phi)= V+ T(\theta,\tilde \phi).
\ee
The corresponding energy density for this kind of shell is ,
\be
\mu=-\frac{1}{32 m^2 \pi} \Big(\nabla^{(2)} T-T+\frac{3 a}{2m}\partial_{\tilde \phi} T \Big),
\ee
which will imply,
\be
\partial_{V} \mu=0.
\ee

\begin{itemize}
\item{No matter on the shell: $S^{ab}=0$}
\par
Given $F(V,\theta,\tilde \phi)= V+ T(\theta,\tilde \phi),$ we also demand $\mu=0.$ This gives,
\be
\nabla^{(2)} T-T+\frac{3 a}{2m}\partial_{\tilde \phi} T =0.
\ee
As we are working in slow rotation regime this equation posses no regular solution like the non rotating case. So we get,
\be 
T(\theta,\tilde \phi)=0.
\ee 
This will make not just the trace part of stress tensor zero but also every component of $\gamma_{ab}$ to be zero. 
\item{Matter without gravitational wave: } Given the supertranslation transformation  we compute the transverse-traceless part of $\gamma_{ab}$,
\be \label{hatgamma}
\hat \gamma_{ab}=\gamma_{ab}-\frac{1}{2}\gamma^{*}g_{ab}+2\gamma_{(a}N_{b)}+(N_{a}N_{b}-\frac{1}{2}N.N g_{ab})\gamma^{\dagger}.
\ee
Now $N^2=0$ and $$\gamma^{*}=\frac{1}{2 m^2\partial_{V} F} \Big(\nabla^{(2)}F-F+V\partial_{V} F  +\frac{3 a}{2m}\partial_{\phi} F\Big), \gamma^{\dagger}=0, \gamma_{a}=\gamma_{ab}n^{b}.$$ We also have, $n^{b}=\partial_{V}$ so only $n^{V}$ component is 1. So we have,
\be
\gamma_{a}=\gamma_{a V}=0.\ee One last ingredient we need is $N_{a}$ and it has only one non vanishing component which is $N_{V}=-1.$ Given this we calculate different components of $\hat \gamma_{ab}$, 
\begin{align}
\begin{split}
&\hat \gamma_{V V}=\gamma_{VV}-2\gamma_{V} +\gamma^{\dagger}=0,\\&
\hat \gamma_{A V}=\gamma_{A V}-\gamma_{A}=0,\\&
\hat \gamma_{\theta\theta}=\gamma_{\theta\theta}-2m^2\, \gamma^{*}=\frac{1}{\partial_{V} F} \Big(2\,\nabla_{\theta}^{(2)} \partial_{\theta} F- \nabla^{(2)} F -\frac{3 a}{2m}\partial_{\phi} F\Big),\\&
\hat \gamma_{\tilde \phi\tilde \phi}=\gamma_{\tilde \phi\tilde \phi}-2 m^2\sin(\theta)^{2}\, \gamma^{*}= \frac{1}{\partial_{V} F}\Big(2\,\nabla_{\tilde\phi}^{(2)} \partial_{\tilde \phi}F-\sin(\theta)^{2}\, \nabla^{(2)} F+\frac{3 a}{2m}\sin(\theta)^{2}\,\partial_{\phi}  F\Big),\\&
\hat \gamma_{\theta\tilde \phi}=\gamma_{\theta\tilde \phi}=\frac{2}{\partial_{V} F}\Big(\nabla_{\theta}^{(2)} \partial_{\tilde \phi} F+\frac{3 a\sin(\theta)^{2}}{2 m}\partial_{\theta} F\Big).
\end{split}
\label{hgamma}
\end{align}
We check that $\hat \gamma_{ab} n^{b}=0, \hat \gamma_{ab}g^{ab}=0.$ Now non existence of gravitational wave should imply $\hat \gamma_{ab}=0.$ One can set the last two equations of (\ref{hgamma}) to be equal to zero and find the relevant soldering freedom. \end{itemize}

\subsection*{\begin{itemize} \item Horizon shell in rotating BTZ \end{itemize}}

The matching conditions are insensitive to the asymptotic structure of the spacetime. In view of this, here we study three dimensional BTZ black hole which is not asymptotically flat, but the supertranslation like soldering freedom still appears. Unlike the Kerr metric, we do not have to consider slow rotation limit for this case. We get an exact analytic answer for arbitrary rotation. So it will be interesting to see whether we get any qualitative changes because of the presence of rotation. The metric takes the following form, 
\be
ds^2= - f(r) dt^2+ \frac{dr^2}{f(r)}+r^2 \Big[N^{\phi}dt +d\phi\Big]^2,
\ee
where, 
$f(r)=-M+(\frac{r}{l})^2+\frac{J^2}{4 r^2}$ and $N^{\phi}(r)=\frac{J}{2}\frac{r^2-r_{h}^2}{r^2 r_{h}^2}.$
Also,
\be
r_{h}^2=\frac{1}{2}\Big(Ml^2+\sqrt{(Ml^2)^2-J^2 l^2}\Big), \tilde r_{h}=\frac{1}{2}\Big(Ml^2-\sqrt{(Ml^2)^2-J^2 l^2}\Big).
\ee
We express all the quantities in terms of $r_{h}$ and $\tilde r_h.$
\begin{align}
\begin{split}
M=\frac{r_{h}^2+\tilde r_{h}^2}{l^2}, \quad J=\frac{2 r_{h} \tilde r_{h}}{l}.
\end{split}
\end{align}
Then we have,
\begin{align}
\begin{split}
N^{\phi}(r)=\frac{\tilde r_{h}}{r_{h}}\frac{r^2-r_{h}^2}{l r^2}\,\quad f(r)=\frac{(r^2-r_{h}^2)(r^2-\tilde r_{h}^2)}{l^2 r^2}.
\end{split}\end{align}
Then we change to Kruskal coordinates \cite{Ross}. 
\be
U=-e^{-\kappa u},\quad V=e^{\kappa v},
\ee
where, $u, v= t\pm r_{*}.$ Also,
\be
r^{*}=\frac{1}{2\kappa}\log\Big(\frac{\sqrt{r^2-\tilde r_{h}^2}-\sqrt{r_{h}^2-\tilde r_{h}^2}}{\sqrt{r^2-\tilde r_{h}^2}+\sqrt{r_{h}^2-\tilde r_{h}^2}}\Big)
\ee
and 
\be
\kappa=\frac{r_{h}^2-\tilde r_{h}^2}{l^2 r_{h}}.
\ee
 Finally we get \cite{Ross},
\be \label{rotBTZ}
ds^2=\frac{1}{(1+ UV)^2}\Big(-4 l^2 dUdV- 4 l \tilde r_{h}( U dV-V dU)d\phi+[(1- UV)^2r_{h}^2+ 4 U V \tilde r_{h}^2]d\phi^2\Big).
\ee
Like the Kerr metric, we again get $Z^{V}=F(V,\phi)$ on the horizon. So there is again an arbitrary soldering freedom in the $V$ direction. Performing an active transformation of the form $V\rightarrow V+F(V,\theta)$ to the  metric of $-$ side  we solder it with usual rotating BTZ metric on $+$ side. Then using (\ref{junc8})  we can write down the components of intrinsic energy momentum tensor of the shell from the jump of the oblique extrinsic curvature.
\begin{align}
\begin{split}
p=-\frac{1}{8\pi}\frac{\partial_{V}^2 F}{\partial_{V} F},J^{\phi}=\frac{1}{4\pi} \frac{ \partial_{V}\partial_{\phi}F}{r_{h}^2\partial_{V}F}, \mu=-\frac{1}{8\pi}\Big(\frac{l^2\partial_{\phi}^2 F-2 l\tilde r_h\partial_{\phi}F-(r_h^2-\tilde r_h^2)(F-V\partial_{V}F) }{r_h^2 l^2 \partial_V F}\Big).
\end{split}
\end{align}
Again setting $p=0$ we get,
\be
F(V,\phi)= V + T(\phi),
\ee
where $T(\phi)$ is an arbitrary function of angular coordinate. 
Also, if we want to set $J^{\phi}=0$ we will get, 
\be
F(V,\phi)= a V+ B(\phi). 
\ee
$B(\phi)$ is an arbitrary function of angular coordinate.  

Also for $p=0$  (or $J^\phi=0)$ case we again retrieve the fact that, $\partial_{V}\mu=0,$
where,
\be
\mu=-\frac{1}{8\pi l^2}\Big(\frac{l^2T''(\phi)-2 l\tilde r_h T'(\phi)-(r_h^2-\tilde r_h^2)T(\phi) }{r_h^2 }\Big).
\ee
The total energy once again is obtained by integrating $\mu$ on the spatial slice of the horizon.
\be
E=\frac{1}{8\pi }\int d\phi \Big(\kappa T(\phi)\Big),
\ee

This matches with the supertranslation charge derived in the literature from different perspectives, for example, readers are referred to \cite{Pinoetal,Barnich:2009se}. Thus as mentioned earlier, BMS-like soldering freedom can also arise for horizon shells placed at spacetimes those are not asymptotically flat.  Next we discuss in detail what happens if we work in Eddington-Finkelstein coordinate. 
 \vskip 0.5 cm
 
 {\bf Eddington-Finkelstein shell:}
 
  To change all the expressions in the Eddington-Finkelstein coordinate, we do the following, 
 \begin{align}
 \begin{split}
F(V,\theta)=e^{\kappa f(v,\theta)}, \partial_{V}=\frac{e^{-\kappa\, v}}{\kappa}\partial_{v},\partial^2_{V}=\frac{e^{-2\kappa\, v}}{\kappa^2}\partial^2_{v}-\frac{e^{-2\kappa\, v}}{\kappa}\partial_{v}.
 \end{split}
 \end{align}
 Using this we get,
\begin{align}
 \begin{split}
&p=-\frac{1}{8\pi}\Big(\frac{\partial_{v}^2 f}{\partial_{v} f}+\kappa\,\partial_{v} f-\kappa\Big),J^{\phi}=\frac{1}{4\pi\,r_h^2}\Big(\frac{\partial_{v}\partial_{\phi}f}{\partial_{v}f}+\kappa\, \partial_{\phi}f\Big),\\&
\mu=-\frac{e^{\kappa \,v}}{8\pi}\Big(\frac{l^2\,\kappa^2 (\partial_{\phi} f)^2+l^2\,\kappa\, \partial_{\phi}^2 f-2 l\tilde r_h\,\kappa\, \partial_{\phi}f-(r_h^2-\tilde r_h^2)(1-\partial_{v}f) }{r_h^2 l^2 \partial_v f}\Big).
\end{split}
\end{align}
We can now study various cases. 
\begin{itemize}
\item{ p=0: This will give,
\be
f(v,\phi)=\frac{1}{\kappa}\Big(\log( e^{\kappa v}+ e^{c_1(\phi)})+c_2(\phi)\Big).
\ee
Translating this into the Kruskal coordinate we get,
\be
F(V,\phi)=e^{c_2(\phi)}V+ T(\phi),
\ee
where, $T(\phi)=e^{c_{2}(\phi)+c_1(\phi)}$.}
\item{ $p=0, J^{\phi}=0:$ This gives, \be e^{c_2(\phi)}=a.\ee $a$ is just a constant. So we get,
\be
F(V,\phi)=a V+ T(\phi).
\ee
  Upto a rescaling factor $a$ we again get supertranslation like transformation. Upto a rescaling factor $a$ we again get supertranslation like transformation. We don't find any regular solution for shell supporting impulsive gravitational waves without any matter for this case also. }

\end{itemize}

\section{Soldering freedom and conformal isometry}
  In this section, we consider various scenarios where special kind of stitching give rise to a new kind of soldering freedom-namely conformal transformations. We have already shown in section (\ref{redux}) that such  situations can arise when the metric on the horizon depends on $V.$ Here we elaborate on this by considering few examples.   
 \begin{itemize}
 \item{{\bf Spacetimes with  constant curvature:} Lets us consider spacetimes with the topology $\mathcal{M}^{1}\times \mathcal{M}^3$.  $\mathcal{M}^{1}$ can be either $R^{1}$ or $S^{1}.$ Similarly  $\mathcal{M}^3$ can be either $R^3$ or $S^3.$ This will enable us to deal with Minkowski, de Sitter and anti-de Sitter spacetimes in a unified way. We will start with 5-dimensional embedding space. 
 \be
- X_0^2+X_1^2+X_2^2+X_3^2+\eta\, X_4^2=\eta\,a^2,\ee
where, $a=\sqrt{\frac{3}{|\Lambda |}}$ and  $\Lambda$ is the cosmological constant. $\eta=1$ for $\Lambda >0,$ $\eta=-1$ for $\Lambda<0$ and $\eta=0$ for $\Lambda=0.$ Then we define \cite{Barrabes1, Hogana, Podolsky},
\begin{align}
\begin{split}
u=\sqrt{2} \,a\,\frac{(X_0+X_1)}{(X_4+a)},v=\sqrt{2}\,a\,\frac{(X_0-X_1)}{(X_4+a)},z=\sqrt{2}\,a\,\frac{(X_2+i\, X_3)}{(X_4+a)}.
\end{split}
\label{emc1}\end{align}
We get,
\be
ds^2=\frac{2 dzd\bar z-2 du dv}{(1+\frac{\Lambda}{6}(z\bar z-uv))^2}.
\label{emb1}\ee
It is evident that this metric represents de Sitter space when $\Lambda >0$, anti-de Sitter space when $\Lambda <0$ and Minkowski space when $\Lambda=0.$  It is well known that the five-dimensional embedding spaces for (anti)de Sitter spacetimes can be parametrized in different ways. Not all of those parametrizations would cover the entire manifold \cite{mal}. However the coordinate systems that we have chosen cover the whole manifold. One can easily see this for Minkowski space where $\eta=\Lambda=0$, and in this case Eq. (\ref{emb1}) produces a Kruskal representation of flat space. Since Kruskal coordinate system is a global set of coordinates ($-\infty < U, V< \infty$), we clearly have a global notion of the light-cone or $U=0$ null hypersurface. The other cases $\Lambda > 0~or~\Lambda<0$ can be viewed similarly \footnote{ Each of the coordinates in (\ref{emc1}) posesses a coordinate singularity at $X_4=-a$.  For more discussions about different choice of embedding coordinates in the context of soldering two spacetimes across a null-shell, interested readers are referred to \cite{Hogana,Podolsky, Podolsky1}.}. Next, we make the following coordinate transformations \cite{Barrabes1, Hogana, Podolsky},
\begin{align}
\begin{split}
u=\frac{z\bar z\, V}{b}-U, v=\frac{V}{b}-\eta\, U,\zeta=\frac{z\, V}{b}, b=1+\eta\,  \zeta\,\bar \zeta .
\end{split}
\end{align}
Then we end up with the following form of the metric,
\be
ds^2=\frac{2 (\frac{V}{b})^2 d\zeta d\bar \zeta +2 dU \, dV-2\,\eta\, dU^2}{(1+\frac{\Lambda\, U}{6}(V-\eta\, U))^2} .
\ee
Now we consider the null surface ($\Sigma$)  $U=0.$ The line element is then simply given by $ds^2|_{\Sigma}=2 (\frac{V}{b})^2 d\zeta d\bar \zeta.$ Recall Eq. (\ref{conf-decomp}), the line elements are almost identical. Next we perform the following infinitesimal conformal transformations,
\be
\frac{2}{b^2} d\zeta d\bar \zeta \rightarrow \frac{2}{b^2} d\zeta d\bar \zeta+\epsilon \frac{2\,\Omega(\zeta,\bar \zeta)}{b^2} d\zeta d\bar \zeta.
\ee
$\epsilon$ is the small parameter and $\Omega(\zeta,\bar \zeta)$ is the conformal factor. 
We then can easily verify that, $$V\rightarrow V(1-\frac{\epsilon\,\Omega(\zeta,\bar \zeta)}{2}),$$ keeps the  the full metric $ds^2|_{\Sigma}$ invariant and  it is  generated by $Z^{V}$ mentioned in  (\ref{mint2}). Also for the individual coordinates we have,
\be
\zeta\rightarrow \zeta+\epsilon\, h(\zeta), \bar \zeta \rightarrow \bar \zeta+ \epsilon\, \bar h(\bar \zeta),\ee
which give,
\be
\Omega(\zeta,\bar \zeta)=\frac{(1+\eta\, \zeta \bar \zeta) \left(h'(\zeta)+\bar h'(\bar \zeta)\right)-2 \,\eta \, \bar \zeta h(\zeta)-2 \eta \, \zeta\, \bar h(\bar \zeta)}{1+\eta \, \zeta \bar \zeta}.
\ee
$h(\zeta)$ and $\bar h(\bar \zeta)$ are holomorphic and anti-holomorphic functions.} These kinds of local conformal transformations are usually referred as superrotation in the context of asymptotic symmetries. There exist infinite way to solder two constant curvature spacetimes. 
Although not exactly in the same context but for horizons described by constant curvature spaces one finds superrotations like transformations while soldering two spacetimes. \\

\item{{\bf Black hole horizon:}  Let us now explore the possibility of recovering such conformal isometries in black hole space times. For four-dimensional black holes (non rotating), the horizon topology is  typically $R\times \mathcal{M}^2$, where $\mathcal{M}^2$ can be $R^2$, $S^2$ or $H^2.$ The metric on $\mathcal{M}^2$ typically takes the form $r^2 d\Omega^2$ where, $r^2$ takes the following form near the horizon in some Kruskal like coordinates \be \label{r} r^2= a+ b\,(U\, V)+ c\, (U\, V)^2+ \cdots.\ee $d\Omega^2$ can either be a metric of 2-dimensional flat, spherical or hyperbolic space. We can immediately see from (\ref{killing2}), at $U=0$ there is no scope of introducing conformal transformations for the spatial cross-section of the horizon. But interestingly if we are not exactly at the horizon but slightly away from it (i.e at $U=\epsilon$, where $\epsilon$ is very small but not zero), then from (\ref{killing2})  and (\ref{mint1a}) we have,
\be 
Z^{V}\,\epsilon\, b\, \tilde g_{AB}+\Omega(x^{A}) (a+ \epsilon\,b\, V )\tilde  g_{AB}=0,
\ee
where $g_{AB}= r^2 \tilde g_{AB}.$ From this we get,
\be \label{conft}
Z^{V}=-\frac{a\,\Omega(x^A) }{b \,\epsilon}-V\, \Omega(x^{A}) +\mathcal{O}(\epsilon).
\ee
We have kept only leading order terms in $\epsilon$. Therefore for a null surface situated close to the black hole horizon, one can have conformal transformations as soldering freedom.  Also it is evident as long as $U\neq 0$, this transformation is non-vanishing but when $U$ becomes zero then we have to set $\Omega(x^{A})$ to zero also.  Hence, $Z^{V}$ will vanish, implying this kind of transformations are not allowed on the event horizon but only permissible outside the event horizon.\par 
 Next we  compute the intrinsic quantities for this type of shell \cite{Blau2}. We will adopt the intrinsic formulation as discussed in Section~(\ref{first}). On the $-$ side we have the usual Schwarzschild metric in Kruskal coordinate as shown below. 
 \be \label{schawz}
ds^2= -G(r) \, dU dV +r^2(U,V)d\Omega_2^2,
\ee
Here $G(r)=\frac{32m^3}{r} e^{-r/2m}$ and $d\Omega_2^2=\frac{dz d\bar z}{(1+z\,\bar z)^2}.$ $z,\bar z$ are the complex coordinates. The shell is located at $U=\epsilon.$ Near the surface we have the following expansion for $r(U,V)$,
 \be
 r(U,V)^2= a+\epsilon\, b\, V,
 \ee 
with, $a= 4m^2, b=-\frac{8 m^2}{e}.$ Here $\epsilon$ is small   and we will only keep terms upto liner order in $\epsilon.$

Now on the $+$ side we take this metric and   perform the following transformation,
\begin{align}
\begin{split}
&V\rightarrow V(1-\tilde \Omega(z,\bar z) )-\frac{a\,\tilde \Omega(z,\bar z) }{b \,\epsilon} +\mathcal{O}(\epsilon),\\&
z\rightarrow z+ h(z),\,\, \bar z\rightarrow \bar z+ \bar h (z).
\end{split}
\end{align}
This induces a conformal transformation on 2-sphere .
\be
d\Omega_2^2= \Big(1+\tilde \Omega (z,\bar z)\Big)d\Omega_2^2,
\ee
with,
\be
\tilde \Omega(z,\bar z)=\frac{(1+z\, \bar z) \left(h'(z)+\bar h'(\bar z)\right)-2 \, \bar z\, h(z)-2  \, z\, \bar h(\bar z)}{1+ z\, \bar z}.
\ee
We only keep  terms  which are only linear in $h(z), \bar h(\bar z)$ or in $\epsilon.$  Using the formulation of \cite{Barrabes, Barrabes1, Poisson} as discussed in Section~(\ref{first}) we compute the jump of the oblique extrinsic curvature of this shell and from that we read off various shell-intrinsic  quantities. Let us discuss the computation briefly. On the side $-$ we will have,
\begin{align}
\begin{split}
n^{\mu}|_{-}=(\partial_{V})^{\mu}, N^{\mu}|_{-}=\frac{1}{G(r(U,V))|_{U=\epsilon}}\, (\partial_{U})^{\mu}.
\end{split}
\end{align}
On the $+$ side we will have,
\begin{align}
\begin{split}
n^{\mu}|_+=(1-\tilde \Omega(z,\bar z))(\partial_{V})^{\mu}.
\end{split}
\end{align}
and the following non vanishing components of the tangent vectors (to the spatial 2 dimensional slice),
\begin{align}
\begin{split}
&e^{V}_{z}|_{+}=\frac{(e-2 V\,\epsilon)\partial_{z}\tilde \Omega(z,\bar z)}{2\,\epsilon}, e^{z}_{z}|_{+}
=1+h'(z),\\&
e^{V}_{\bar z}|_{+}=\frac{(e-2 V\,\epsilon)\partial_{\bar z}\tilde \Omega(z,\bar z)}{2\,\epsilon}, e^{\bar z}_{\bar z}|_{+}=1+\bar h'(\bar z).
\end{split}
\end{align}
Auxiliary normal $ N^{\mu}|_{+}$ can be easily determined uniquely by solving the following constraints,
\be
N^{\mu}\, n_{\mu}=-1, N^2=0, N_{\mu}\, e^{\mu}_{z}=N_{\mu}\, e^{\mu}_{\bar z}=0.
\ee 
Now using (\ref{junc8}) we get the following non-vanishing quantities, 
\begin{align}
\begin{split}
&J^{z}=\frac{ (1+z \bar z)^2\,\partial_{\bar z} \tilde \Omega(z,\bar z) }{64\,m^2\, \pi}+\mathcal{O}(\epsilon),J^{\bar z}= \frac{ (1+z \bar z)^2\,\partial_{ z} \tilde \Omega(z,\bar z) }{64\, m^2\,\pi}+\mathcal{O}(\epsilon),\\&\mu=\frac{(e+  V \, \epsilon ) \Omega (z,\bar z)}{16 \pi  m^2 \, \epsilon}+\frac{V( \bar z\, h(z)+ z \bar h(\bar z))}{16 \, m^2\, \pi  \,  (1+ z \bar z)}+\mathcal{O}(\epsilon).
\end{split}
\end{align}
} \end{itemize}
Here again, we have kept only the leading order terms. The pressure of the shell is zero but there are non-zero currents and energy density. Due to the explicit dependence of  the energy density $\mu$ on $V$, this shell satisfies a conservation equation as shown in (\ref{consv}).


\section{Summary and Discussion}
 We have revisited the dynamics of thin null shell situated at the horizon of black holes in general relativity and explored the freedom of patching two metrics across such horizon shells. We have analysed horizon shells for rotating black holes in four and three dimensions. If we stitch two rotating metrics (slow rotation for Kerr metric) across the shell demanding the induced metric remains invariant up to its isometric transformations, we recover BMS-like transformations.  We computed the shell stress-energy tensor from the jump of the oblique extrinsic curvature of the shell. 
 For spacetime with different asymptotic structure like BTZ we also produce BMS-like soldering symmetries. This emphasizes the fact that the appearance of soldering symmetries is a local phenomenon. It will be interesting to find this for the Kerr shell without considering the slow rotation limit as many features of axisymmetry get suppressed under the assumption of slow rotation. In that case, we probably get a non-trivial shell with different intrinsic properties.  
 
 Role of conformal invariance in the near horizon physics of black hole has been explored quite extensively in recent past \cite{Brown-Hennaux,Strominger:1997eq, Carlip:1998wz,Carlip:1999cy}. Much of the recent activities related to BMS algebra are reconsideration of asymptotic symmetries in relation to flat space holography \cite{Arcioni:2003xx, Arcioni:2003td,Barnich:2010eb}; that is, extending the AdS/CFT  correspondence for asymptotically flat spacetimes. The emergence of superrotation as a new symmetry of the near horizon physics is an important outcome of these line of research \cite{Barnich:2009se,Compere,Pinoetal,Strominger}. Superrotation type symmetries may also be recovered from the soldering perspective for certain kind of horizon shells. Especially space times with constant curvature exhibit such kind of soldering freedom. In some special coordinate systems, the soldering group in such spacetimes (in 4 dimensions) becomes a combination of shift in some suitable null direction (for example advanced null coordinate $v$) and infinite dimensional conformal isometries on 2-surface. In this note, we have deduced a generic condition when this will occur and illustrated prominent examples of this fact. We have shown, if the induced metric on the null hypersurface can be expressed as a product of some function of lightlike direction and spherical metric (or any 2-d spatial metric), we can allow conformal isometries together with some shift in the degenerate null direction to solder the spacetimes across the surface. This freedom seems to be absent for black holes, as in general they do not admit such expressibility of induce metric. However, near horizon shell of a black hole can admit such freedom. This may have interesting consequences upon the near horizon degrees of freedom of black holes. 
 \par

We anticipate that the results obtained here can have far reaching consequences specially in the context of black hole information paradox.  It will be interesting to investigate the soldering symmetries for time dependent shells. At late time we expect the emerging symmetry will be close to the BMS like soldering transformations \cite{Tanabe}. We hope to get back to this issue in near future. Also it would be interesting to study any possible connection between black hole membrane paradigm and supertranslation-superrotation transformations and how the BMS like soldering transformations affect the properties of the horizon fluid \cite{Penna:2015gza,Eling}.

 \section*{Acknowledgements}
Authors are grateful to Matthias Blau for many insightful comments and important suggestions which have helped to improve this draft. Authors would like to thank Amitabh Virmani, Avirup Ghosh, Swastik Bhattacharya, and Sudipta Sarkar, and Alok Laddha for their valuable comments. AB would like to thank Prof. Ling-Yan Hung for discussions and constant encouragements and  acknowledges the support from the Thousand Young Talents Program, and Fudan University. SB acknowledges the warm hospitality of Institute for Theoretical Physics, University of Bern where part of this work was done. The work of SB is supported by SERB Early Career Research Award ECR/2017/002124.

\appendix

\bibliographystyle{JHEP}

\end{document}